\documentclass{Interspeech2024}
\usepackage{float}
\usepackage{balance}
\usepackage{verbatim}

\usepackage{tikz}
\usepackage{amsmath}
\usetikzlibrary{shapes.geometric, arrows, positioning}

\tikzstyle{process} = [rectangle, minimum width=3cm, minimum height=1cm, text centered, draw=black]
\tikzstyle{datastore} = [cylinder, draw, shape border rotate=90, minimum height=2cm, minimum width=1cm, text centered]
\tikzstyle{arrow} = [thick,->,>=stealth]
\tikzstyle{privacy} = [text=blue]
\tikzstyle{bias} = [text=red]

\interspeechcameraready

\title{Examining the Interplay Between Privacy and Fairness for Speech Processing: A Review and Perspective}

\name[affiliation={1}]{Anna}{Leschanowsky}
\name[affiliation={2}]{Sneha}{Das}

\address{
  $^1$Fraunhofer Institute for Integrated Circuits IIS, Germany\\
  $^2$Dept. of Applied Mathematics and Computer Science, Technical University of Denmark, Denmark}
\email{anna.leschanowsky@iis.fraunhofer.de, sned@dtu.dk}


\begin{document}

\maketitle

\begin{abstract}

    Speech technology has been increasingly deployed in various areas of daily life including sensitive domains such as healthcare and law enforcement. For these technologies to be effective, they must work reliably for all users while preserving individual privacy. Although
    tradeoffs between privacy and utility, as well as fairness and utility, have been extensively researched, the specific interplay between privacy and fairness in speech processing remains underexplored. This review and position paper offers an overview of emerging privacy-fairness tradeoffs throughout the entire machine learning lifecycle for speech processing. By drawing on well-established frameworks on fairness and privacy, we examine existing biases and sources of privacy harm that coexist during the development of speech processing models. We then highlight how corresponding privacy-enhancing technologies have the potential to inadvertently increase these biases and how bias mitigation strategies may conversely reduce privacy. By raising open questions, we advocate for a comprehensive evaluation of privacy-fairness tradeoffs for speech technology and the development of privacy-enhancing and fairness-aware algorithms in this domain. 

\end{abstract}

\section{Introduction}

Privacy and fairness are often treated as separate domains with unique challenges and solutions. Traditionally, advancements in one area have been pursued independently of the other which has led to isolated silos in the research community. However, recent research has shown that privacy-enhancing technologies can influence bias in machine learning algorithms~\cite{shaham2023holistic, khalili2021improving, pannekoek2021investigating, sarhan2020fairness}. On the other hand, fairness-aware model learning techniques have shown adverse effects on an individual's privacy~\cite{shaham2023holistic, chang2021privacy, zhang2023interaction}. While few studies have started investigating this tradeoff and model training techniques that can ensure fairness, privacy and accuracy at the same time~\cite{zhang2023interaction}, the nature of privacy-fairness tradeoff within speech processing is yet to be explored. 

This article challenges the assumption that privacy and fairness can be addressed separately in the context of speech processing technologies and highlights the need for exploring privacy-fairness tradeoffs in this domain. Given that modern speech processing technologies rely heavily on machine learning and generative models, the privacy-fairness tradeoffs observed in other domains are likely applicable to speech processing technology. Therefore, we highlight that privacy and fairness are interdependent dimensions that should be addressed together. Enhancing one aspect, whether through bias mitigation or privacy-enhancing technologies, can inadvertently diminish the other. This suggests that strategies effective in one domain may have unintended consequences in the other leading to privacy-fairness tradeoffs.  

To allow investigations of this tradeoff, it is essential to understand the specific privacy harms and biases present in machine learning-based speech processing models. Activities that can lead to privacy harm can range from surveillance during data collection to disclosure of personal data after model deployment. Similarly, biases can manifest all throughout the machine learning (ML) life cycle, during data collection, model building and model evaluation, leading to disparities in subgroup performance. By recognizing these harms as well as bias mitigation and privacy-enhancing techniques (PETs) throughout the ML life cycle, we lay the foundation towards harmonized strategies that address both privacy and fairness. 
The main contributions of this semi-review-perspective paper are: 
\begin{itemize}
    \item RQ1 Which privacy harms and bias coexist in the machine learning life cycle in the context of speech processing?
    \item RQ2 How do privacy-fairness trade-offs occur in ML-based speech processing models? 
    \item RQ3 What are the open questions on privacy-fairness trade-offs for speech processing technology?
\end{itemize}

\section{Background and Related Work}

\subsection{Bias and Fairness}
Algorithmic fairness deals with the detection, quantification and mitigation of bias in decision-making systems~\cite{barocas2017fairness}. AI fairness 360 (AIF360), Fairlearn, Aequitas are among the tools that aid in addressing bias and ensuring {\it algorithmic} fairness in systems~\cite{bellamy2019ai, bird2020fairlearn, saleiro2018aequitas}. While there is no universally agreed upon definition or understanding of fairness~\cite{saxena2019perceptions}, most technical work refer to one or more of individual, group and sub-group fairness~\cite{kearns2018preventing}. Equalised odds, equal opportunity, demographic parity, treatment equality, test and counterfactual fairness are commonly employed fairness metrics~\cite{mehrabi2021survey}. Incorporating fairness through the above metrics can happen at various stages of model development. During pre-processing, data can be transformed to remove underlying discrimination with respect to protected groups. By incorporating fairness metrics within the objective or loss function, models can be trained to abide by the desired fairness metric. 

Within recent years, a large body of research has been developed within algorithmic fairness for computer vision and natural language processing (NLP). Much of this advancement can also be applied to spoken language. However, a recent investigation on NLP tasks shows that language and communication itself can inherit differential handling of groups due to its intrinsic connection to society~\cite{baugh2000racial, loudermilk2015implicit, craft2020language}. In other words, when working with language, discrimination can be encoded more implicitly within the data~\cite{blodgett-etal-2020-language}. Therefore, {\it techniques for bias mitigation and the evaluation of systems with respect to fairness should be reassessed in the context of the application}, to move beyond groups and individuals.

In addition to the technical challenges of developing fair algorithmic systems, there is a growing discourse on {\it if and when} fairness can be automated. Much of the work on fairness centers around the US legal framework. Work addressing the mapping of fairness technologies to the EU legal framework has begun only recently. One of the earliest papers on the topic~\cite{wachter2021fairness} highlights the gap between automated fairness and EU non-discrimination law. Furthermore, there is a disconnect between statistical measures of fairness and contextual sensitivity. This is due to the difference between EU legal framework, retained as largely agile to address context-based discrimination. Besides, statistical metrics for fairness can sometimes seem to be conflicting and require considerations of the context within which the AI is applied \cite{binns2020apparent}. Quoting from \cite{wachter2021fairness}, {\it while numerous statistical metrics exist in technical literature, none can yet reliably capture a European conceptualisation of discrimination.} 
This highlights the need to not only look at bias and fairness in conjunction with the target spoken language task the model is applied in, but also the intersection of algorithmic fairness with the legal, ethical and political aspects of fairness in speech processing. 

\subsection{Privacy and Privacy Harms}

While various notions of privacy have been discussed in the literature, there is no singular, universally accepted definition of privacy~\cite{solove2005, nautsch2019gdpr}. This has led legal scholars to suggest shifting from the vague concept of ``privacy" to more concrete activities that can cause privacy harms~\cite{solove2005}. Consequently, these activities can result in privacy violations. For instance, the act of disseminating information by publishing a person's speech recording without their consent constitutes a privacy violation. In consequence, the individual might experience privacy harm because of unwanted disclosure. One of the earliest identified privacy harms is dignitary harm, e.g., reputation injury~\cite{warren1989right}. Other privacy harms are related to activities that increase the risk of future dignitary, monetary or physical harm to an individual~\cite{solove2005}. In the context of speech processing, the collection of a person's voice recordings could lead to future identity theft resulting in dignitary and monetary harm. Privacy harms also include activities that create societal or institutional power imbalances~\cite{solove2005}. For example, the development and use of speaker recognition systems by law enforcement can disrupt power balances and lead to abuses. 

Importantly, activities can lead to multiple privacy harms simultaneously such as the aggregation of personal information which can result in both dignitary harm and power imbalances, as one party acquires far more knowledge about an individual than expected. Privacy-enhancing technologies play a crucial role in addressing these issues, as they restrict activities that could result in privacy problems. For instance, anonymizing data helps prevent the identification or disclosure of private information, thereby minimizing the risk of identity theft and dignitary harm. Thus, by preventing potentially harmful activities, PETs can preserve and strengthen an individual's privacy.



\subsection{Privacy-Fairness Trade-Off in Other Domains}

While the tradeoffs between privacy and utility and fairness and utility have been extensively discussed~\cite{srivastava2022privacy, nautsch2019preserving, gu2023elucidate}, there remains a gap in understanding the interplay between privacy and fairness specifically in the speech processing field. Existing research in other domains has shown conflicting findings regarding the relationship between privacy and fairness~\cite{shaham2023holistic}. Some studies suggest that enhancing fairness can have positive implications for privacy~\cite{dwork2012fairness, fiorettodifferential, aalmoes2022leveraging}, while other research indicates tradeoffs where efforts to achieve fairness inadvertently compromise privacy for certain subgroups~\cite{chang2021privacy, zhang2023interaction}. Similarly, privacy-preserving methods have been shown to both positively and negatively impact fairness measures during model training~\cite{khalili2021improving, pannekoek2021investigating, sarhan2020fairness}. A recent study on the unfairness of privacy-enhancing technologies has addressed privacy-fairness tradeoffs from a legal and computer science point of view~\cite{calvi2024unfair}. They discuss possible technical as well as regulatory solutions including the usage of data protection impact assessment for evaluating PETs and enabling context-specific decision-making while considering their limitations and impact on fairness. So far, studies on privacy-fairness tradeoffs have been conducted in the field of natural language processing~\cite{matzken-etal-2023-trade, lyu-etal-2020-differentially, maheshwari-etal-2022-fair} and on a variety of classification tasks~\cite{bagdasaryan2019differential, Pujol2020, farrand2020}. However, investigations on privacy-fairness tradeoffs in speech processing tasks are yet to be explored. 

Moreover, most of the previously discussed studies have predominantly investigated techniques applied during model training, e.g. differential privacy or dropout-based debiasing techniques, while overlooking privacy-fairness tradeoffs across the entire machine learning life cycle. In the context of datasets, one study has discussed fairness, privacy preservation and regulatory compliance within biometric datasets~\cite{mittal2023responsible}. Their investigations, however neglecting voice-based datasets, highlight a fairness-privacy paradox as sensitive attribute information enables fairness quantification but compromises privacy~\cite{mittal2023responsible}. In the voice domain, a qualitative analysis of voice biometric datasets shed light on bias and privacy challenges but did not quantify the fairness-privacy paradox~\cite{rusti2023benchmark}. 
Therefore, there remains the need to explore privacy-fairness tradeoffs comprehensively from data collection to model deployment, particularly in the domain of speech processing. 


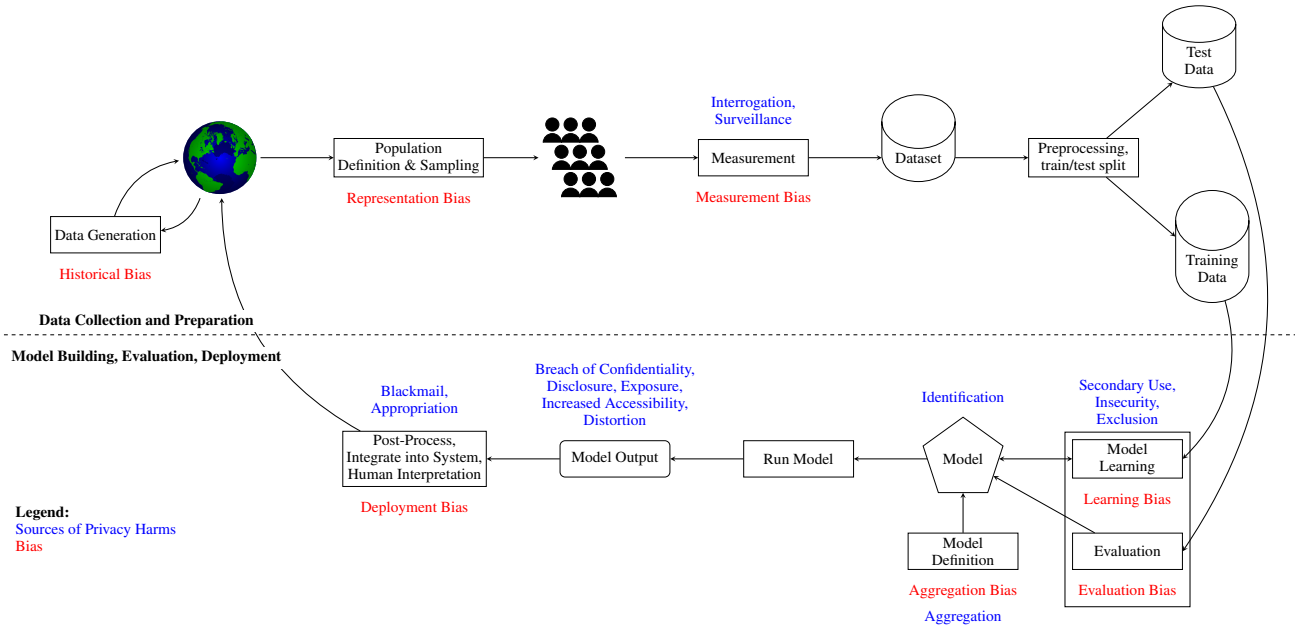
\begin{figure*}[!htb]
  \centering

\usetikzlibrary{shapes.geometric, arrows, positioning, fit}

\tikzstyle{process} = [rectangle, minimum width=3cm, minimum height=1cm, text centered, draw=black]
\tikzstyle{output} = [rectangle, minimum width=3cm, minimum height=1cm, text centered, draw=black, rounded corners]
\tikzstyle{datastore} = [cylinder, draw, shape border rotate=90, minimum height=2cm, minimum width=2cm, text centered]
\tikzstyle{cube} = [regular polygon, regular polygon=6, minimum width=2cm, text centered, draw=black]
\tikzstyle{arrow} = [thick,->,>=stealth]
\tikzstyle{privacy} = [text=blue]
\tikzstyle{bias} = [text=red]


    \resizebox{\textwidth}{!}{%
        \begin{tikzpicture}[node distance=1.5cm, every node/.style={fill=white, font=\large}, align=center]

        \node (globe) [draw=none, fill=none] {\includegraphics[width=2cm]{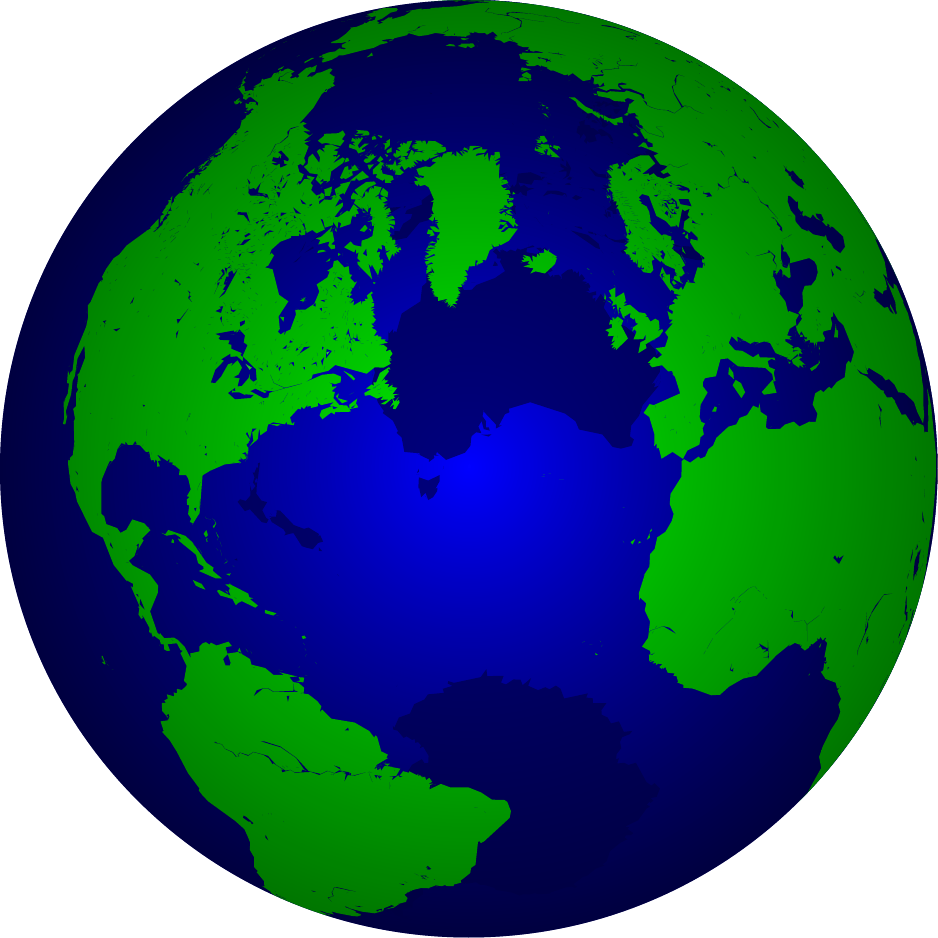}};
        
        \node (data_generation) [process, below left=0.5cm and 0.5cm of globe] {Data Generation};
        \node (population) [process, right=2cm of globe] {Population\\Definition \& Sampling};
        \node (humans) [draw=none, fill=none, right=1.5cm of population] {
            \begin{tikzpicture}
                \foreach \i in {0, 1, 2} {
                    \foreach \j in {0, 1, 2} {
                        \pgfmathsetmacro{\x}{\j * 0.5 - \i * 0.25}
                        \pgfmathsetmacro{\y}{\i * 0.75}
                        
                        \draw[fill=black] (\x, \y + 0.6) circle [radius=0.18cm];

                        \draw[thick,fill=black] (\x + 0.25,\y + 0.1) arc[start angle=0,end angle=180,radius=0.25cm];
                    }
                }
            \end{tikzpicture}
        };
        \node (measurement) [process, right=2cm of humans] {Measurement};
        \node (dataset) [datastore, right=2cm of measurement] {Dataset};
        \node (preprocessing) [process, right=2cm of dataset] {Preprocessing,\\train/test split};
        \node (testdata) [datastore, above right=1.5cm and 1cm of preprocessing] {Test\\Data};
        \node (trainingdata) [datastore, below right=1.5cm and 1cm of preprocessing] {Training\\Data};
        \node (modellearning) [process, below left=4cm and 0.1cm of trainingdata] {Model\\Learning};
        \node (evaluation) [process, below=1.5cm of modellearning] {Evaluation};
        \node (model) [cube, left=2cm of modellearning] {Model};
        \node (modeldefinition) [process, below=1.1cm of model] {Model\\Definition};
        \node (runmodel) [process, left=2cm of model] {Run Model};
        \node (modeloutput) [output, left=2cm of runmodel] {Model Output};
        \node (postprocess) [process, left=2cm of modeloutput] {Post-Process,\\Integrate into System,\\Human Interpretation};

        \draw [arrow] (globe) to[in=90, bend left](data_generation);
        \draw [arrow] (data_generation) to[out=270, bend left] (globe);
        \draw [arrow] (population) -- (humans);
        \draw [arrow] (humans) -- (measurement);
        \draw [arrow] (measurement) -- (dataset);
        \draw [arrow] (dataset) -- (preprocessing);
        \draw [arrow] (preprocessing) -- (testdata);
        \draw [arrow] (preprocessing) -- (trainingdata);
        \draw [arrow] (trainingdata) to[out=90, in=270, bend left] (modellearning.east);
        \draw [arrow] (testdata) to[out=90, in=90, bend left] (evaluation.east);
        \draw [arrow] (evaluation) -- (model);
        \draw [arrow, <->] (modellearning) -- (model);
        \draw [arrow] (model) -- (runmodel);
        \draw [arrow] (runmodel) -- (modeloutput);
        \draw [arrow] (modeloutput) -- (postprocess);
        \draw [arrow] (modeldefinition) -- (model);
        
        \draw [arrow] (globe) -- (population);

        \node[privacy, above=0.3cm of measurement, align=center] {Interrogation,\\Surveillance};
        \node[bias, below=0.3cm of population, align=center] {Representation Bias};
        \node[bias, below=0.3cm of measurement, align=center] {Measurement Bias};
        \node (historicalbias) [bias, below=0.3cm of data_generation, align=center] {Historical Bias};
        \node[bias, below=0.3cm of modellearning, align=center] {Learning Bias};
        \node[privacy, above=0.3cm of modellearning, align=center] {Secondary Use, \\Insecurity, \\Exclusion};
        \node (evaluationbias)[bias, below=0.3cm of evaluation, align=center] {Evaluation Bias};
        \node[privacy, above=0.3cm of model, align=center] {Identification};
        \node[bias, below=0.3cm of modeldefinition, align=center] {Aggregation Bias};
        \node[privacy, below=1cm of modeldefinition, align=center] {Aggregation};
        \node[privacy, above=0.3cm of modeloutput, align=center] {Breach of Confidentiality,\\Disclosure, Exposure,\\Increased Accessibility,\\ Distortion};
        \node[privacy, above=0.3cm of postprocess, align=center] {Blackmail,\\ Appropriation};
        \node[bias, below=0.3cm of postprocess, align=center] {Deployment Bias};

        \node[draw=black, fill=none, thick, fit={(modellearning) (evaluation) (evaluationbias)} ,inner sep=0.2cm] (box) {};

        \node[draw=none, fill=none, below=6cm of historicalbias, align=left] (legend) {
            \begin{tabular}{ll}
                \textbf{Legend:} & \\
                \textcolor{blue}{Sources of Privacy Harms} \\ 
                \textcolor{red}{Bias}
            \end{tabular}
        };

        \draw [arrow, color=black] (postprocess) to[bend left] (globe);

        \draw[dashed, thick] ([yshift=-0.5cm]current bounding box.west) -- ([yshift=-0.5cm]current bounding box.east);
        
        \node at (-2, -4.5) {\textbf{Data Collection and Preparation}};
        \node at (-2, -5.5) {\textbf{Model Building, Evaluation, Deployment}};

        \end{tikzpicture}%
    }


  \caption{We depict the general ML lifecycle including associated biases and sources of privacy harms (adapted from~\cite{suresh2021} and~\cite{solove2005}).}
  \label{fig:privacybias}
\end{figure*}

\section{Sources of Privacy Harms and Bias in Speech Processing Tasks}
\label{sec:PrivacyHarmsBias}

We draw on two well-established frameworks in the privacy and fairness field to show the coexistence of privacy harms and biases in the machine learning (ML) life cycle~\cite{suresh2021, solove2005}. While many different types of bias have been categorized within the ML community~\cite{mehrabi2021survey, Olteanu2019social}, the framework by~\cite{suresh2021} identifies seven potential sources of harm throughout the ML life cycle. They distinguish types of bias within data collection and preparation and model development, evaluation, postprocessing and deployment. Particularly, their framework has been proven useful to assess bias in speaker recognition models~\cite{hutiri2022bias}. To identify sources of privacy harm, we draw on a well-established taxonomy that maps out activities that can cause privacy harm~\cite{solove2005}. It distinguishes between four types of harmful activities namely information collection, information processing, information dissemination and invasion. While this taxonomy has been proposed before recent advances in AI, it is sufficiently broad and technology-agnostic to be used in the context of smart home devices~\cite{chhetri2022mute} and AI technologies~\cite{Lee2024}. Moreover, activities of information collection, processing and dissemination can be easily mapped to the ML life cycle. We do not claim to provide an exhaustive analysis of harmful activities but rely on \textcolor{black}{this taxonomy} as a baseline to understand sources of privacy harms alongside sources of bias. 

In the following sections, we present sources of privacy harms and bias in more depth including examples from speech processing tasks. This serves as a foundation to discuss potential privacy-fairness tradeoffs in the speech processing domain. Figure~\ref{fig:privacybias} provides an overview of privacy risks and biases throughout the entire machine learning lifecycle.


\subsection{Data Collection and Preparation} 
In this section, we describe the types of biases and sources of privacy harms that can arise during speech data collection and preparation. 

\subsubsection{Sources of bias during data curation} 
\noindent\textbf{Historical Bias:} This type of harm refers to the reflection of existing societal biases in datasets~\cite{suresh2021}. For instance, the VoxCeleb 1 dataset, commonly used for speaker verification tasks, has been shown to exhibit historical bias~\cite{hutiri2022bias}. Moreover, word embeddings widely used for natural language processing have been shown to reflect real-world biases~\cite{suresh2021}. This is likely to extend to speech processing technology with the implementation of large language models for ASR error correction~\cite{schmidt2024generative, chen2024hyporadise}.

\noindent\textbf{Representation Bias:} Representation bias occurs when a subset of the population is underrepresented. This can result from the underrepresentation of certain groups in the sample, a mismatch between the target population and the user population or limitations in the sampling method~\cite{suresh2021}. Representation bias has been explored within the speech domain and has been identified in datasets used for developing speaker recognition systems~\cite{hutiri2022bias, rusti2023benchmark, fenu2021fair, fenu2020improving} and ASR systems~\cite{garnerin2019gender, maison23_interspeech}. Other work has explored the impact of representation bias on model performance related to gender, content and prosody for various downstream tasks, e.g., phoneme recognition, keyword spotting, emotion recognition~\cite{meng2022don}. 

\noindent\textbf{Measurement Bias:} This type of bias is related to the collection of features or labels used as proxies for prediction. Issues with such proxies can range from oversimplification to measurement variations across groups, such as differences in method and accuracy~\cite{suresh2021}. For example, the VoxCeleb dataset uses nationality as subgroup labels conflating nationality with accent and dialect~\cite{hutiri2022bias}.

\subsubsection{Sources of privacy harm during data curation}
\noindent\textbf{Surveillance:} Audio surveillance has long been regarded as problematic. However, the rise of speech assistants and their potential to eavesdrop on private conversations, surveillance has become a significant privacy threat~\cite{edu2020smart}. The issue is particularly concerning when these recordings are reviewed by humans for the purpose of training automatic speech recognition models~\footnote{\url{https://www.bbc.com/news/technology-47893082}}. Surveillance has been shown to create discomfort, may result in self-censorship and can be a barrier for the adoption of speech technology~\cite{lau2018alexa, tabassum2019investigating}. 

\noindent\textbf{Interrogation:} Interrogation can take on many forms with varying degrees of coercion. While interrogation has been traditionally understood as directly pressuring individuals to disclose information, pressure can also be indirect or subtle~\cite{solove2005}. Individuals may feel interrogated and pressured even if they have previously consented to data collection. For instance, speech corpora like Mixer come with rich metadata including family history and smoking status potentially having led participants to perceive interrogation~\cite{rusti2023benchmark}.
In addition, \cite{solove2005} argues that interrogation can lead to data distortion due to the power and control of the interrogator which can result in measurement bias. As speech technologies allow users to socially and naturally engage with a system, risks of interrogation and unwanted extraction of information from the user increase~\cite{Lee2024}. 

\subsection{Model Building, Evaluation and Post-Processing}

We now describe types of biases and sources of privacy harms during model building, evaluation and post-processing. While information processing risks to privacy have been discussed more generally for AI technology~\cite{Lee2024}, we map them to data processing in the context of training and building machine learning models for speech processing tasks. Thereby, we do not focus on how machine learning models can enable harmful activities, e.g. speaker identification models enable automated identification at scale, but how data processing to build machine learning models can cause privacy harms, e.g. identification through identity inference attacks. 

\subsubsection{Sources of bias during model building and evaluation}

{\noindent\bf Learning Bias:} This type of bias arises if model choices affect performance disparities across samples~\cite{suresh2021}. For instance, changes in model size can affect models for speaker recognition~\cite{hutiri2022bias, fenu2020improving} and keyword spotting~\cite{hutiri2023tiny}. Similarly, design decisions in the development of voice anonymization systems have been shown to influence subgroup performance~\cite{leschanowsky2024voice}. 

{\noindent \bf Aggregation Bias:} Aggregation bias occurs when a general model does not adequately fit data consisting of underlying groups~\cite{suresh2021}. It can be related to representation bias if the model primarily fits the over-represented group in the dataset. However, aggregation bias arises during model building rather than dataset collection as it stems from the decision to aim for a generalizable model. Aggregation bias has been frequently explored in various speech processing tasks by comparing group performance in ASR~\cite{tatman2017gender, tatman17_interspeech, koenecke2020racial, feng2021quantifying}, speaker verification systems~\cite{hutiri2022bias, fenu2021fair, chen2022exploring} and speech emotion recognition~\cite{slaughter2023pre}.

{\noindent \bf Evaluation Bias:} Evaluation bias occurs when the evaluation dataset does not match the use population due to representation bias of the benchmark dataset or the use of singular metrics that can obscure aggregation bias~\cite{suresh2021}. 

\subsubsection{Privacy harm during model building and evaluation}
{\noindent \bf Aggregation:} Aggregation involves combining data to reveal detailed information about a person~\cite{solove2005}. This process occurs during information processing and does not require the acquisition of new data but rather the use of already collected data. While speech alone is a rich source of information, its combination with other data sources can significantly enhance insights. For instance, cross-modal fusion techniques that rely on audio, as well as corresponding transcripts, have been explored for emotion recognition~\cite{sebastian2019fusion}. Multimodal approaches have also been applied in the health domain for depression assessment showing superior performance compared to uni-modal frameworks~\cite{zhao2022unaligned}. 

{\noindent \bf Identification:} Identification poses a significant privacy risk by enabling the linkage of data to a specific individual~\cite{solove2005}. It is important to distinguish between models designed to identify or verify a person's identity based on biometric markers, such as their voice, and models developed for other tasks that can still identify individuals. Generally, speaker identification differs from speaker verification which has benefits such as accessing various accounts, e.g. bank account, while reducing fraud and increasing usability~\cite{campbell1997speaker}. However, since voice is a biometric marker, any speech processing model that records a person's voice can potentially leak identity information. Foremost, training data for speech processing models contains a person's identity regardless of whether it is necessary for the task. This makes it exploitable by attackers~\cite{han2020voice}. Moreover, identity inference attacks enable attackers to link a victim's recording to their identity~\cite{xiao2023micpro, feng2023review}. Additionally,~\cite{srivastava2019privacy} showed that ASR-encoded representation carries enough information to identify a person. 

{\noindent \bf Insecurity:} Insecurity is an increasing risk associated with the harmful activities of data aggregation and identification~\cite{solove2005}. When aggregated or identifiable information is stored, the potential consequences of insecurity and subsequent harm can be more severe. While the insecure storage of datasets is critical, insecure processing of data for model building and deployment is equally concerning. Furthermore, speech processing models typically rely on centralized training paradigms which are vulnerable to cybersecurity attacks~\cite{feng2023review}. Alternative training methods, such as federated learning, have been explored to enhance data security by keeping training data secure and only transmitting model parameters. Yet, federated learning cannot be considered a fully privacy-preserving method as the transmission of model parameters remains susceptible to attacks~\cite{feng2023review}.

{\noindent \bf Secondary Use:} Secondary use refers to the use of data for purposes other than those originally intended~\cite{solove2005}. This practice is a significant privacy concern, especially for speech processing tasks which generally rely on the collection of speech data. Collected speech data to train an ASR model can easily be reused for other speech-related tasks. For instance, the Mixer corpora, initially collected for the development of speaker recognition models, have been used to train models on smoking status identification due to their rich metadata~\cite{ma2022automatic}. Secondary use poses dignitary harm as people might provide or donate their speech data for a specific purpose and would not consent to its use for other purposes~\cite{solove2005}. To address this, configurable privacy-preserving voice processing has attempted to disentangle voice signals to ensure that only authorized tasks can be performed~\cite{aloufi2020privacy}.

{\noindent \bf Exclusion:} Exclusion is related to the lack of transparency and the inability of individuals to exercise control over their data~\cite{solove2005}. Exclusion can be considered privacy harm during model building if individuals are not informed that their speech data is being used to train speech-centric models. For instance, the VoxCeleb dataset has been scraped from YouTube to enable speaker verification in-the-wild~\cite{nagrani2020voxceleb}. To provide end-users with notice and control, speech assistants provide privacy settings where users can permit the use of their speech recordings for training purposes~\footnote{\url{https://www.amazon.co.uk/gp/help/customer/display.html?nodeId=GVP69FUJ48X9DK8V}}. 

\subsection{Model Deployment}

Finally, we provide an overview of sources of privacy harms and biases when deploying a model in a real-world setting. While information dissemination risks are generally associated with revealing or sharing personal information~\cite{solove2005}, we do not consider risks associated with data sharing~\cite{Lee2024}. Instead, we focus on harmful activities that can be enabled by only deploying a model.

\noindent\textbf{Deployment Bias: } Deployment bias arises when the originally defined problem space of the model does not match the actual usage~\cite{suresh2021}. For instance, speaker verification models used in the forensic domain will have different requirements than those used as proof-of-life systems for pensioners which need to be considered during model deployment~\cite{hutiri2022bias}. Moreover, outcomes of speech processing models used in the medical context, e.g., depression detection, are often interpreted by human decision-makers who can be subject to automation or confirmation bias. 

\subsubsection{Sources of privacy harm during deployment}
\noindent\textbf{Breach of Confidentiality, Disclosure and Exposure:} Breach of confidentiality is a privacy harm that appears independent of data that has been revealed but dependent on whether a breach of trust has happened~\cite{solove2005}. If a person trusts a company with their speech data for model training, they do not expect their data to be disclosed. In contrast, disclosure involves revealing true information about a person to others causing harm primarily through reputational damage rather than a breach of trust~\cite{solove2005}. Finally, exposure involves exposing certain physical and emotional attributes about a person, which can lead to embarrassment and humiliation~\cite{solove2005}. It is strongly related to societal norms and violation of these norms can result in intense feelings of shame.

In the context of model deployment, both breaches of confidentiality and disclosure can occur due to the revelation of sensitive attributes resulting from model attacks. These attacks can expose sensitive information or membership to the training dataset, leading to both a breach of confidentiality and potential disclosure. For instance, speech recognition models~\cite{shah2021evaluating, miao2021audio} and speaker recognition systems~\cite{chen2023slmia} have been shown to be vulnerable to membership inference attacks. Additionally, property inference attacks can reveal gender information when applied to speech emotion recognition tasks~\cite{feng2021attribute}. If such attacks are used on speech-centric models related to medical conditions, individuals can be easily exposed. 

\noindent\textbf{Increased Accessibility: } Increased accessibility enhances risk of disclosure-related harms~\cite{solove2005}. Here, the harm does not arise from directly disclosing personal information but from making it more easily accessible. Although deploying a model does not directly reveal information about an individual, it can lead to the disclosure of sensitive information through model attacks. Therefore, deploying a model can be seen as increasing accessibility to information about an individual. 

\noindent\textbf{Blackmail:} While breach of confidentiality, disclosure and exposure are considered privacy harms due to the actual revelation of information, blackmail involves the threat of disclosure~\cite{solove2005}. If deployed models are attacked and personal information about an individual is gathered, the attacker could use this information to blackmail the individual. Furthermore, speech generation technology can exacerbate risks in creating fake but convincing content for blackmailing~\cite{Lee2024, Hutiri2024}.

\noindent\textbf{Appropriation:} Using another person's identity or personality is considered a privacy violation regardless of whether the appropriation is disrespectful~\cite{solove2005}. Recent advancements in speech synthesis make appropriation a growing privacy concern. With the development of such models, it has become easier than ever to impersonate an individual using only a few seconds of audio~\cite{Hutiri2024}. For instance, a branch manager in Hong Kong paid \$35 Million to fraudsters after being scammed by fake audio~\footnote{\url{https://www.forbes.com/sites/thomasbrewster/2021/10/14/huge-bank-fraud-uses-deep-fake-voice-tech-to-steal-millions/}}. Moreover, identity spoofing attacks enable an attacker to exploit an individual's voice recording for purposes such as speaker recognition~\cite{xiao2023micpro}.

\noindent\textbf{Distortion:} While disclosure involves revealing true information about an individual, distortion relates to spreading false information that can lead to reputational harm and embarrassment~\cite{solove2005}. Current speech synthesis models can generate realistic audio that can result in economic, social and reputational harms~\cite{hutiri2022bias}. For instance, voices of celebrities have been cloned to share violent and harmful content~\footnote{\url{https://www.vice.com/en/article/dy7mww/ai-voice-firm-4chan-celebrity-voices-emma-watson-joe-rogan-elevenlabs}}. In addition, several incidents have been reported where the voice of deceased people has been generated using state-of-the-art speech generation~\cite{Hutiri2024}.

\section{Contextualizing Privacy-Fairness Tradeoffs for Speech Processing}
\label{sec:Privacy-FairnessTradeoff}

In the previous section, we identified various privacy harms and biases present at different stages of the machine learning life cycle and discussed them in the context of speech processing. These harms inevitably coexist when developing models for various speech-processing tasks. 
In the following section, we discuss the influence of bias mitigation strategies on privacy and the effect of PETs on fairness, with a particular focus on speech processing.

\subsection{PETs and their Impact on Fairness}

{\bf Anonymization:} Anonymization refers to the process of removing identifiers from data to prevent the re-identification of individuals. \textit{Anonymization can eliminate privacy harms during model building as well as during model deployment}. Yet, anonymization techniques can hinder bias detection by removing sensitive attributes essential for identifying bias~\cite{calvi2024unfair}. Additionally, de-anonymization attacks can be more successful against protected groups~\cite{ekstrand2018privacy, calvi2024unfair}. 
In speech processing, anonymization must address both the linguistic content of the signal and the speaker's voice. These systems, however, can suffer from representation bias, learning bias and aggregation bias as training datasets and model choices can impact subgroup performance~\cite{leschanowsky2024voice, zhu2023investigating}. 
While these studies have identified group biases, other work has investigated individual differences among speakers in voice anonymization~\cite{williams2024anonymizing, sinhaeli}. 
Content anonymization, on the other hand, focuses on removing sensitive semantic information that would allow re-identification such as names, addresses or bank account numbers~\cite{williams22_spsc}. Previous work on speech content privacy has utilized keyword recognition to remove or replace sensitive content~\cite{qian2018towards, qian2019speech, hu2022speechhide} or end-to-end speech transcription for dummy word injections~\cite{ahmed2020preech}.  

\noindent
{\bf Synthetic Data:} Synthetic data is fully or partially generated artificially while preserving properties of the original data~\cite{calvi2024unfair}. \textit{Synthetic data can mitigate privacy harms during all stages of the ML lifecycle} and has been recently emphasized in the EU AI Act. However, despite advances in synthesizing speech data, the generation process depends on original data, which may be biased, leading to the replication of these biases in the synthetic data~\cite{calvi2024unfair}. 
Synthetic data - generated using text-to-speech systems - has been utilized to augment existing datasets for training speech recognition models~\cite{rosenberg2019speech, hu2022synt++} and keyword spotting models~\cite{sharma2020adaptation, lin2020training, werchniak2021exploring} but its impact on biases has not been explicitly explored. Other work has employed anonymization techniques to create synthetic data based on the VoxCeleb dataset but found similar performance disparities as in the original dataset hinting at replication biases~\cite{miao2024synvox2}.


\noindent
{\bf Differential Privacy:} Differential privacy can be used as a privacy-enhancing technique by adding random noise to a dataset while preserving its statistically significant insights~\cite{calvi2024unfair}. Intuitively, differential privacy enables an individual to contribute to a dataset without increasing privacy risks. Differentially private models build on this idea in that no additional information can be learned from any individual training sample~\cite{nautsch2019preserving}. Therefore, \textit{differential privacy can mostly restrict privacy harms during model deployment} as it prevents models from leaking information about training data. However, its application during model building can impact related biases such as aggregation or learning bias.  
Few papers have applied differential privacy to speech processing focusing on creating privacy-preserving speech data releases~\cite{han2020voice}, replacing sensitive keywords in output transcripts~\cite{ahmed2020preech} and developing differentially private speaker anonymization~\cite{shamsabadi2023differentially, hu2022speechhide, Qian2018HideBehind}, speech recognition~\cite{yang2022ensemble} and speech emotion recognition models~\cite{feng2022user}.

\noindent
{\bf \textcolor{black}{Cryptographic Methods:}} \textcolor{black}{Cryptographic methods can encompass homomorphic encryption, secure multiparty computation and distance-preserving hashing techniques~\cite{nautsch2019preserving}.}
Homomorphic encryption allows usage, processing and computations on encrypted data and can be divided into three types, i.e. partially, somewhat and fully homomorphic encryption~\cite{nautsch2019preserving}. \textcolor{black}{Secure multiparty computation enables multiple parties to compute a function while preserving the privacy of their inputs, and distance-preserving hashing techniques can similarly satisfy privacy constraints with minimal computational overhead~\cite{pathak2012privacy}.} \textit{In general, cryptographic methods can prevent privacy harms during model building and deployment} by increasing the security of the model and decreasing risks related to confidentiality and disclosure. Homomorphic encryption has been used to securely train speech processing models like keyword spotting systems~\cite{zheng2022keyword, elworth2022hekws}, speech emotion recognition~\cite{dias2018exploring}, and health-related paralinguistic tasks~\cite{teixeira2018patient}. \textcolor{black}{Secure multiparty computation and hashing techniques have been used for speaker recognition and identification as well as speech recognition~\cite{pathak2012privacy}.} As encryption is a reversible process and does not alter data, its influence on bias and fairness is limited. Nevertheless, applying homomorphic encryption limits the operational and architectural choices of the model which can influence learning bias~\cite{franco2021toward}.  

\noindent
{\bf Federated Learning:} Federated Learning allows training of machine learning models on multiple devices while transferring updated parameters to the server~\cite{feng2023review}. As it does not rely on transferring speech data to a central server, \textit{federated learning has the potential to not only mitigate privacy risks during model development and deployment but also to reduce privacy risks related to data collection}. Federated learning has been explored for speech recognition~\cite{yufederated, zhu22b_interspeech}, speech emotion recognition~\cite{tsouvalas2022privacy}, keyword spotting~\cite{leroy2019federated} and speaker verification~\cite{granqvist20_interspeech}. However, it is important to note that federated learning is not fully privacy-preserving as transmitted parameters can leak sensitive information and therefore, has to be explored in combination with differential privacy~\cite{feng2023review, feng2022user, shoemate2022sotto}. Federated learning has been shown to influence fairness due to general underlying biases, party selection and the propagation of bias in other domains~\cite{chang2022bias, abay2020mitigating}.

\subsection{Bias, Fairness and their Impact on Privacy}
We discuss the impact of fairness on privacy by focusing on two prerequisites of bias assessment and mitigation: a) sensitive-attribute consolidation, b) model-fairness. 

\noindent
{\bf Attribute-specific Data Collection: }Sensitive attributes play an important role in the assessment and detection of bias in ML algorithms \cite{andrus2022demographic}, and the subsequent mitigation strategies to ensure fair algorithms. Within the realms of ML and speech processing, complimenting core data with potentially sensitive attributes such as age, gender, accent, or health conditions is crucial for ensuring fairness and mitigating biases in speech models. These attributes are used to identify and correct biases that lead to unfair treatment of different user groups, such as mis-detection of speech from individuals with accents or from outside the normative age group or population \cite{feng2021quantifying, das2022speech}. Without the additional attribute information,  models might perform well on average while systematically underperforming for the underrepresented, but this performance disparity may go unnoticed. Incorporating sensitive attributes into datasets is thus vital for creating more inclusive and equitable speech processing systems.

 The inclusion of sensitive attributes in speech datasets can lead to {\it all} the privacy harms depicted in Fig.~\ref{fig:privacybias}, from surveillance to identification and appropriation. Therefore, attribute-specific data collection and processing raises technical, ethical and legal concerns and can be in conflict with data minimization requirements. Research has shown that attacks can be staged to infer personal or non-target attributes of individuals or the context from the trained models \cite{parisot2021property, ganju2018property}. Furthermore, regulations such as the GDPR impose strict guidelines on the handling of sensitive data, including the necessity for explicit consent and the right to data erasure \cite{andrus2022demographic}. In addition, the EU AI-act and the Digital Services Act (DSA) require fairness guarantees for AI models. These regulations aim to protect individuals' privacy and prevent data from being used in ways that could harm them. Balancing the need for sensitive data to achieve fairness with the obligation to protect privacy requires careful consideration of data minimization principles, ensuring that only the necessary data is collected and that it is used transparently and ethically. 

\noindent
{\bf Fair Models: }In addition to the privacy harms arising due to the consolidation of sensitive data, fairness metrics can also impact privacy. The authors in \cite{chang2021privacy} investigated the influence of {\it fairness criterion - equalized odds} on the success of membership inference attacks. This criterion was applied during in-processing and post-processing stages of the ML-lifecycle. The paper shows that fairness constraints widen the difference in privacy risks among the subgroups, with the underprivileged groups facing higher privacy risks in terms of successful membership inference attacks. This could be due to the lack of generalizability of models from underrepresented groups thereby leading to to memorization of training instances and potential information leakage, which can lead to privacy harms like identification and appropriation. So far, these observations have been made on simulated data and the COMPAS and law datasets. {\it Resampling and reweighting} based bias mitigation strategies could be prone to a similar trade-off. On an algorithmic level, a paper employing graph neural networks shows that the privacy risks on the edge increase when individual fairness at the nodes is improved \cite{zhang2024unraveling}.

\section{Open Questions on Privacy-Fairness Tradeoff for Speech Processing Technology}

In Section~\ref{sec:PrivacyHarmsBias} and~\ref{sec:Privacy-FairnessTradeoff} we discussed biases, sources of privacy harms and associated mitigation strategies in the context of speech processing. By understanding PETs and bias mitigation strategies in relation to the harms that they address, we can now point out where and how privacy-fairness tradeoffs are likely to occur within the ML lifecycle in the context of speech processing. Moreover, our framework helps to outline which biases and privacy risks need to be evaluated in combination and which PETs and bias mitigation strategies can potentially be combined to create privacy-enhancing and fairness-aware strategies. 
In the following parts, we raise 
high-level questions and potential solutions that we urge the speech-processing community to explore in the future. 

\noindent\textbf{How prevalent are privacy-fairness tradeoffs for speech processing technologies?}

\noindent
Based on our review of PETs and bias mitigation strategies in Section~\ref{sec:Privacy-FairnessTradeoff}, there's limited research on privacy-fairness tradeoffs for speech processing technologies. 
To the best of our knowledge, this study presents the first framework that contextualizes the privacy-fairness tradeoff for speech processing 
by exploring the theoretical interplay between privacy risks, PETs, and bias and fairness strategies through the ML lifecycle. We demonstrated that both differential privacy and homomorphic encryption can mitigate sources of privacy harm during model building and deployment. In addition, we find that certain biases, such as aggregation and learning bias, arise during this stage and might be impacted by these technologies. For example, homomorphic encryption restricts operational and architectural choices raising questions about how these restrictions influence the learning bias of models. By using the framework to connect harms, PETs and bias mitigation strategies, one can develop a meaningful evaluation plan to investigate the prevalence of the privacy-fairness tradeoff in various speech processing tasks through the ML-lifecycle. 
Future research could build on this work to balance privacy and fairness in speech-processing technologies.

\noindent
\textbf{What should be considered when addressing privacy-fairness tradeoffs (during speech-data collection)?} 

\noindent
We find that historical, representation and measurement bias coexist with potential privacy risks such as interrogation and surveillance during data collection and preparation. During the data collection stage, bias assessment relies on the collection of sensitive attributes. While anonymization might be necessary for dataset publishing, it can hinder bias detection based on these attributes. Therefore, anonymization must be applied thoughtfully to allow for required bias evaluations. Balancing the need for sensitive data to achieve fairness with the obligation to protect privacy requires careful consideration of data minimization principles, ensuring that only the necessary data is collected and used transparently and ethically. As long as bias detection relies on sensitive attributes, it will be necessary to assess the associated risks of PETs on fairness in datasets and to consider the trade-offs. 

PETs that can mitigate privacy risks during data collection include synthetic data or federated learning as they do not require speech data to be collected for the speech-based ML model under development. Recent advances in synthetic speech data generation enable the use of synthetic data to address privacy and ethical challenges. The field of face recognition has seen increased usage and research into diverse and fair synthetic face datasets and similar trends could be expected in speech processing. However, synthetic data generated using text-to-speech synthesis can suffer from biases due to bias propagation, similar to synthetic face datasets~\cite{huber2024bias}. In particular, research on synthetic datasets needs to avoid ``diversity-washing" and ``consent circumvention" risks~\cite{whitney2024real}. Federated learning can potentially mitigate privacy risks but has been shown to influence fairness due to inherent biases, party selection and bias propagation~\cite{chang2022bias, abay2020mitigating}. Here, biases in federated learning for speech processing are yet to be explored.

\section{Conclusion}
Privacy, bias and fairness are critical topics in the current and timely discourse on safety in artificial intelligence. Through this review-position paper, we emphasize the need to understand privacy and fairness as mutually dependent concepts, particularly for speech communication due to the biometric and sensitive nature of speech signals. Motivated by existing frameworks on privacy and fairness, in this paper, we contextualize a {\it privacy-fairness framework for ML-based speech processing}. We show how sources of privacy harm and bias coexist at every stage of the ML lifecycle. Additionally, we present existing research that investigates the impact of a) PETs on bias, and b) fairness on privacy. While the former is relatively well understood in the ML community, it is yet to be quantified for speech technologies, and studies on the impact of fairness on privacy are generally limited. These insights inform our final contribution which includes open questions and potential solutions to address the privacy-fairness tradeoff in speech models.

\balance
\bibliographystyle{IEEEtran}

\bibliography{mybib}

\end{document}